%% file: main.tex
\documentclass[%
 reprint,
 amsmath,amssymb,
 aps,
]{revtex4-2}

\usepackage{graphicx}
\usepackage{dcolumn}
\usepackage{bm}
\usepackage{physics}
\usepackage{caption}
\usepackage{subcaption}

\begin{document}

\preprint{APS/123-QED}

\title{Analytic spectrum of multi-frequency Raman generation with chirped pulses}

\author{Joscelyn van der Veen}
\email{joscelyn.vanderveen@utoronto.ca}
\affiliation{Department of Physics, University of Toronto, Toronto, Canada}


\date{\today}

\begin{abstract}
Multi-frequency Raman generation is a promising method of producing ultrashort laser pulses with high intensities and visible wavelength frequencies. In experimental realizations of multi-frequency Raman generation the Raman order peaks display several behaviours that are not explained by the current theory. We derive an analytic and perturbative expression for the spectrum radiated by a Raman medium excited by multiple Gaussian laser pulses and show that it qualitatively agrees with experimental observations.
\end{abstract}

\maketitle


\section{Introduction \label{sec:intro}}
\input{intro}

\section{Two-State Schrödinger Equation \label{sec:two-state-SE}}
\input{two-state-SE}

\section{Amplitudes for States of Interest \label{sec:amplitudes}}
\input{amplitudes}

\section{Polarization \label{sec:polarization}}
\input{polarization}

\section{Spectrum \label{sec:spectrum}}
\input{spectrum}

\section{Discussion \label{sec:discussion}}
\input{discussion.tex}

\section{Conclusion \label{sec:conclusion}}
\input{conclusion}

\begin{acknowledgments}
We wish to acknowledge Daniel James, Donna Strickland, Stephen Rand, Zujun Xu, and Kaleb Ruscitti for useful correspondence and discussion. We also acknowledge the support of the Natural Sciences and Engineering Research Council of Canada (NSERC).
\end{acknowledgments}

\appendix

\input{appendix}

\bibliography{references}

\end{document}

%% file: intro.tex
Ultrafast lasers are useful for a variety of applications such as medicine \cite{Svanberg2004} and spectroscopy \cite{Maiuri2019}. These applications make use of the femtosecond or sub-femtosecond timescales of ultrashort pulses to image with molecular precision or access the powers required for nonlinear effects. There are currently processes that can generate ultrashort pulses at a wide range of wavelengths but they have low peak intensities due to the inefficiency of the processes \cite{Maiuri2019, Boyd}. Multi-frequency Raman generation (MRG) is a promising method of generating ultrashort laser pulses with high intensities at a range of visible frequencies \cite{Yan2013}. 

When light interacts with a medium, it normally exhibits spontaneous Raman scattering with a small proportion of the incoming light being scattered at a frequency offset by the Raman frequency of the medium. However, when the medium is instead excited by two lasers: a pump laser and a probe laser that is offset from the pump by the Raman frequency, there is efficient conversion of energy from the pump beam frequency to the probe beam frequency, which is called stimulated Raman scattering \cite{Boyd}. Energy can also be efficiently converted to higher Raman order frequencies: frequencies which are offset from the pump frequency by multiples of the Raman frequency. This effect is known as multi-frequency or high-order Raman generation \cite{Wilke1978}. Due to this high efficiency, MRG can produce a spectrum of discrete Raman order peaks that have both short durations (on the order of femtoseconds) and high intensities. The efficient conversion process was first described theoretically by Hickman et al. \cite{Hickman1986}.

Experimentally, there are other optical effects that can occur and compete with the MRG process for energy. For example, a process known as self-phase modulation can compete with MRG but this can be avoided by using chirped pulses \cite{Sali2004}. However, when exciting the Raman medium with chirped pulses, the Raman orders have a sideband shifted to a lower frequency \cite{Turner2008}. These red-shifted shoulders on the Raman orders are not predicted by the current theory. Further, the central frequency of the Raman orders depends on the energy of the pump and probe beams, red-shifting as the energy increases \cite{Yan2013}. 

To understand the process that generates the red-shifted shoulders and find ways to incorporate them into future theory, we explicitly determine an analytic expression for the spectrum of radiation emitted by the Raman medium. A preliminary report of these results was given in \cite{vanderVeen:22, PN2022}

%% file: two-state-SE.tex
We first review the theory of a Raman medium excited by a multiwave field, which is described in Hickman et al. \cite{Hickman1986}.

In absence of an electric field, the Raman medium has the Hamiltonian $H_0$ with eigenfunctions $\ket{n}$ and eigenenergies $\hbar W_n$. The electric field of the Raman orders perturbs $H_0$ with potential $V(z, t)=-p\epsilon(z, t)$ where $p$ is the electric dipole moment operator and $\epsilon(z, t)$ is the electric field. Here we have assumed the electric field propagates spatially in the z-direction. To describe the propagation of the multiple Raman order fields, we take the electric field to be a sum of the orders,

\begin{equation}
    \epsilon(z, t)=\frac{1}{2}\sum_j\epsilon_j(z, t)e^{i\omega_jt}+c.c.
    \label{eq:electric-field}
\end{equation}

\noindent where $\epsilon_j(z, t)$ is the amplitude of the electric field of the $j$th Raman order with frequency $\omega_j=\omega_0+j\omega_R$. We denote the pump laser frequency as $\omega_0$ and the Raman frequency as $\omega_R$.

We wish to obtain the solution to the Schrödinger equation

\begin{equation}
    \left(H_0+V(z,t)\right)\ket{\psi(z, t)}=i\hbar\frac{\partial}{\partial t}\ket{\psi(z,t)}.
    \label{eq:SE}
\end{equation}

To find this solution, we begin by writing the wavefunction in terms of the unperturbed energies. 

\begin{equation}
    \ket{\psi(z,t)}=\sum_{n=1}^\infty c_n(z, t)e^{-iW_nt}\ket{n}
    \label{eq:psi}
\end{equation}

The amplitudes $c_n(z,t)$ can be found from our perturbed Schrödinger equation and will determine the wavefunction of the Raman material under excitation by the multiwave electric field.

Since we are considering here the reaction of the Raman medium to the electric field at some specific position, we will no longer explicitly write the dependence on $z$. With this notation, when we evaluate Eq. \ref{eq:SE} with $\ket{\psi}$ given by Eq. \ref{eq:psi}, we obtain the set of coupled equations:

\begin{multline}
    i\hbar\frac{\partial c_n}{\partial t} = -\frac{1}{2}\sum_{n'}\sum_jc_{n'}(t)p_{nn'}\left[\epsilon_j(t)e^{i(\omega_j+W_{nn'})t}\right. \\
    +\left. \epsilon_j^*(t)e^{-i(\omega_j-W_{nn'})t}\right]
    \label{eq:dcn-dt}
\end{multline}

\noindent where $W_{nn'}\equiv W_n-W_{n'}$ and $p_{nn'}\equiv \mel{n}{p}{n'}$.

Since this is a Raman process, the medium is excited from a vibrational state to a high energy virtual state and then decays to another vibrational state. To do MRG, we excite the medium with two lasers separated by approximately the difference in the ground and first vibrational states. We call this difference the Raman frequency of the medium. We assume that transitions between the ground and first vibrational states do not occur directly and the only transitions are between these two states and higher vibrational states. Mathematically, we assume the only nonzero dipole moment terms are $p_{1n}=p_{n1}$ and $p_{2n}=p_{n2}$ where $n\neq1,2$. We can therefore simplify Eq. \ref{eq:dcn-dt} for the higher electronic states, $n>2$, in terms of just the amplitudes $c_1(t)$ and $c_2(t)$.

It is now important to consider the regime of the laser pulse duration compared to the frequency of electric field oscillations. We want to compare this theory to experiments with visible range frequencies and pulses that are chirped to long time durations (on the order of hundreds of femtoseconds or longer) while propagating through the Raman medium. We are therefore in the adiabatic regime where the pulse is effectively constant over an oscillation of the electric field.

In this regime, the amplitudes $c_n(t)$ and pulse shapes $\epsilon_j(t)$ are both slowly varying in time compared to the oscillations of the complex exponentials. With this assumption there are several ways of integrating Eq. \ref{eq:dcn-dt}, the simplest being to hold $c_{n'}$ and $\epsilon_j$ constant while integrating the exponential. 

We can then write a simplified Schrödinger equation for the ground ($c_1(t)$) and first ($c_2(t)$) vibrational states.

\begin{equation}
    i\hbar\frac{\partial}{\partial t} 
    \begin{pmatrix} c_1(t) \\ c_2(t) \\ \end{pmatrix} = 
    \begin{pmatrix} H_{11}(t) & H_{12}(t) \\ H_{21}(t) & H_{22}(t) \\ \end{pmatrix}
    \begin{pmatrix} c_1(t) \\ c_2(t) \\ \end{pmatrix}
    \label{eq:two-state-SE}
\end{equation}

\noindent where 

\begin{align}
    \begin{split}
        H_{11}(t)&=-\frac{\alpha_{11}}{4}\sum_j\epsilon_j(t)\epsilon_j^*(t) \\
        H_{12}(t)&=-\frac{\alpha_{12}}{4}\sum_j\epsilon_j(t)\epsilon_{j-1}^*(t) \\
        H_{21}(t)&=H_{12}^*(t) \\
        H_{22}(t)&=-\frac{\alpha_{22}}{4}\sum_j\epsilon_j(t)\epsilon_{j}^*(t) \\
    \end{split}
    \label{eq:hamiltonian}
\end{align}

\noindent and

\begin{equation}
    \alpha_{ik}(\omega_j)=\frac{1}{\hbar}\sum_{n=3}^\infty p_{in}p_{nk}\left(\frac{1}{W_{ni}-\omega_j}+\frac{1}{W_{nk}+\omega_j}\right).
    \label{eq:polarizability}
\end{equation}

We call $\alpha_{ik}$ a two-photon polarizability since it describes the effect of an electric field on a state with some intermediate state. We have assumed $W_{21}\approx\omega_R$ to write the two-photon polarizability. We will assume that $\alpha_{ik}$ are constant for the rest of the calculations since $\omega_j$ are far from resonance.

The two-state Schrödinger equation is the same as derived in Hickman et al. \cite{Hickman1986}

%% file: amplitudes.tex
To find an analytic approximation for the emitted radiation, we now deviate from the usual theory and determine the amplitudes for the states of interest, $c_1(t)$ and $c_2(t)$, by solving the two-state Shrödinger equation (Eq. \ref{eq:two-state-SE}) directly. Since the Hamiltonian is time-dependent, we expand it perturbatively in a Dyson series. We can determine the order of the perturbation from the degree of the two-photon polarizability $\alpha_{ik}$, which are all small parameters.

We assume the system begins in the ground state so our initial condition is,

\begin{equation}
    \begin{pmatrix} c_1(-\infty) \\ c_2(-\infty) \\ \end{pmatrix} = \begin{pmatrix} 1 \\ 0 \\ \end{pmatrix}.
    \label{eq:initial-conditions}
\end{equation}

We now need to make an assumption about the general shape of the electric field amplitudes $\epsilon_j(t)$. We choose a linearly chirped Gaussian pulse shape to resemble the experimental laser pulses.

\begin{equation}
    \epsilon_j(t)=A_je^{-\beta_jt^2/4}
    \label{eq:V-of-time}
\end{equation}

\noindent where we have a complex Gaussian width $\beta_j=\frac{1}{a_j^2}-ib_j$ dependent on the pulse width $a_j$ and the linear chirp $b_j$. We assume the coefficient $A_j=E_je^{i\phi_j}$ is time-independent. We also note that taking the complex conjugate of $\epsilon_j(t)$ will only change the time-dependence of the field by changing the sign of the chirp $\beta_j^*=\frac{1}{a_j^2}+ib_j$.

Assuming a chirped Gaussian pulse shape, the amplitudes $c_1(t)$ and $c_2(t)$ are straightforward to calculate to first order. For the second order term, we must calculate double integrals of the form, 

\begin{equation}
    D(a, b; t)=\int_{-\infty}^t\int_{-\infty}^{t'} \exp\left(-\frac{t'^2}{a^2}\right)\exp\left(-\frac{t''^2}{b^2}\right)\dd{t''}\dd{t'}
    \label{eq:second-order-general}
\end{equation}

where $a, b$ are each one of $2/\sqrt{\beta_j+\beta_{j-1}^*}$ and $\sqrt{2}a_j$. Since the pulse widths of the Raman orders are similar, we can assume $|a-b|$ is small.

Integrating over $t''$, we are left with the integral of a Gaussian function and the integral of a Gaussian function multiplied by an error function. Since $|a-b|$ is small, we will write $b=a+\delta$ in the error function and Taylor expand it to first order about $\delta=0$.

We then have the general form for the second order term of the amplitudes,

\begin{equation}
    D(a, b; t) = \frac{\pi a b}{8}\left(1+\erf\left(\frac{t}{a}\right)\right)^2 + \frac{\delta b}{4}\exp\left(-\frac{2t^2}{a^2}\right) + \mathcal{O}(\delta^2)
    \label{eq:second-order-solved}
\end{equation}

Thus, the amplitudes for the ground and first vibrational state are to second order,

\begin{subequations}
\begin{multline}
    c_1(t) = 1 + \frac{i\alpha_{11}}{4\hbar}\sum_j|A_j|^2\sqrt{\frac{\pi}{2}}a_j\left(1+\erf\left(\frac{t}{\sqrt{2}a_j}\right)\right) \\
    +\frac{\alpha_{11}^2}{16\hbar^2}\sum_j\sum_k|A_j|^2|A_k|^2 D\left(\sqrt{2}a_j, \sqrt{2}a_k; t\right) \\
    + \frac{\alpha_{12}^2}{16\hbar^2}\sum_j\sum_kA_jA_{j-1}^*A_k^*A_{k-1} \\
    \times D\left(\frac{2}{\sqrt{\beta_j+\beta_{j-1}^*}},\frac{2}{\sqrt{\beta_k^*+\beta_{k-1}}};t\right)
    \label{eq:c1-first-order}
\end{multline}
\begin{multline}
    c_2(t) = \frac{i\alpha_{12}}{4\hbar}\sum_j A_j^* A_{j-1} \sqrt{\frac{\pi}{\beta_j^*+\beta_{j-1}}} \\
    \times\left(1+\erf\left(\frac{\sqrt{\beta_j^*+\beta_{j-1}}t}{2}\right)\right) \\
    + \frac{\alpha_{11}\alpha_{12}}{16\hbar^2}\sum_j\sum_kA_j^*A_{j-1}|A_k|^2D\left(\frac{2}{\sqrt{\beta_j^*+\beta_{j-1}}},\sqrt{2}a_k;t\right) \\
    + \frac{\alpha_{12}\alpha_{22}}{16\hbar^2}\sum_j\sum_k|A_j|^2A_k^*A_{k-1}D\left(\sqrt{2}a_j,\frac{2}{\sqrt{\beta_k^*+\beta_{k-1}}};t\right).
    \label{eq:c2-first-order}
\end{multline}
\label{eq:amp-first-order}
\end{subequations}

%% file: polarization.tex
The polarization induced by the sum of Gaussian electric fields is given by the expectation value of the dipole moment.

\begin{equation}
    \expval{p}=\mel{\psi(t)}{p}{\psi(t)}
    \label{eq:exp-dipole-moment}
\end{equation}

To determine the polarization, we substitute Eq. \ref{eq:psi} for $\ket{\psi(t)}$ and then take the Fourier transform. We define the Fourier transform as,

\begin{equation}
    \hat{f}(\omega)\equiv\mathcal{F}\{f(t)\}(\omega)=\int_{-\infty}^\infty f(t)e^{-i\omega t}\dd{t}
    \label{eq:fourier-transform}
\end{equation}

\noindent and the inverse Fourier transform as,

\begin{equation}
    f(t)\equiv\mathcal{F}^{-1}\{\hat{f}(\omega)\}(t)=\frac{1}{2\pi}\int_{-\infty}^\infty \hat{f}(\omega)e^{i\omega t}\dd{\omega}.
    \label{eq:inv-fourier-transform}
\end{equation}

If we recall our assumption that $p_{1n'}$ and $p_{n2}$ are the only nonzero dipole matrix elements, we can rewrite the polarization as a single infinite sum over $n>2$. We can also consider Eq. \ref{eq:dcn-dt} written in terms of just the amplitudes $c_1(t)$ and $c_2(t)$ for $n>2$. By taking the Fourier transform of Eq. \ref{eq:dcn-dt} for $n>2$, substituting it into our polarization, and recalling our assumption that $W_{21}\approx\omega_R$, we can write the Fourier transform of the polarization solely in terms of convolutions of $\hat{c}_1$, $\hat{c}_2$, and $\hat{\epsilon}_j$.

The Fourier transform of the polarization is given by the sum,

\begin{multline}
    \mathcal{F}\{\expval{p}\}=\sum_{n=3}^\infty\sum_j\sum_{p=1,2}\sum_{q=1,2}p_{pn}p_{nq} \\
    \times\left(\hat{S}_{pq}(\omega,\omega_{j+p-q},\epsilon_j)+\hat{S}_{pq}(\omega,-\omega_{j-p+q},\epsilon_j^*)\right)+c.c.
    \label{F-dipole-moment-sum}
\end{multline}

over terms of the following form,

\begin{multline}
    \hat{S}_{pq}(\omega,\omega',\epsilon_j) = \frac{-1}{8\pi^2\hbar}\int_{-\infty}^\infty \hat{c_p^*}(\omega-u-W_{pn})\frac{1}{u} \\
    \times\int_{-\infty}^\infty\hat{c}_q(y)\hat{\epsilon}_j(u-y+W_{pn}-\omega') \dd{u}\dd{y}
    \label{eq:spectrum-term}
\end{multline}

To integrate this general term, we first determine the Fourier transform of the general amplitude $c_p(t)$. This general amplitude is a sum over the different time-dependencies of the amplitudes given by Eq. \ref{eq:amp-first-order}. 

\begin{multline}
    c_p(t) = \eta_p^{(0)} + \sum_j \left[\eta_{pa}^{(1)}(j)\erf\left(\frac{t}{\sqrt{2}a_j}\right)\right. \\ 
    + \eta_{pb}^{(1)}(j)\erf\left(\frac{\sqrt{\beta_j+\beta_{j-1}^*}t}{2}\right) + \eta_{pa}^{(2)}(j)\erf\left(\frac{t}{\sqrt{2}a_j}\right)^2 \\
    + \eta_{pb}^{(2)}(j)\erf\left(\frac{\sqrt{\beta_j+\beta_{j-1}^*}t}{2}\right)^2+\eta_{pa}^{(3)}(j)e^{-t^2/a_j^2} \\
     + \left.\eta_{pb}^{(3)}(j)e^{-(\beta_j+\beta_{j-1}^*)t^2/2}\right]
    \label{eq:c1-of-t-simplified}
\end{multline}

The Fourier transform of the error function is $\mathcal{F}\{\erf(t)\}=-2ie^{-\omega^2/4}/\omega$. We can determine the Fourier transform of the square of the error function by taking a convolution of two error functions. The resulting integrals are improper but we will take the principal value since they describe a physical system.

Using partial fraction decomposition, we can separate the integrand into two terms with a single pole each. We can then identify the principle value of the Fourier transform of $\erf^2(t)$ as a sum of Hilbert transforms of Gaussian functions, where the Hilbert transform is defined as,

\begin{equation}
    H(y)=\frac{1}{\pi}P.V.\int_{-\infty}^\infty\frac{e^{-x^2}}{y-x}\dd{x}
    \label{eq:hilbert-transform}
\end{equation}

\noindent and it can be shown that the Hilbert transform of a Gaussian function is proportional to the Dawson function $F(x)$ (see Appendix \ref{app:dawson}). 

Using the oddness of the Dawson function, we find that,

\begin{equation}
    \mathcal{F}\{\erf^2(t)\}=\frac{-8e^{-\omega^2/8}}{\sqrt{\pi}\omega}F\left(\frac{\omega}{2\sqrt{2}}\right).
    \label{eq:F-of-erf-squared}
\end{equation}

We thereby determine $\hat{c}_p(\omega)$ and then can integrate Eq. \ref{eq:spectrum-term} over $y$. Since $\hat{\epsilon}_j$ is a Gaussian function, we have to integrate several more terms that are Hilbert transforms of Gaussian functions. We also have to integrate Hilbert transforms of Gaussian functions that are multiplied by Dawson functions.

Consider, for example,

\begin{multline}
    \int_{-\infty}^\infty e^{-(u-y+W_{pn}-\omega')^2/\beta_j}\frac{e^{-a_l^2y^2/4}}{y}F\left(\frac{a_ly}{2}\right)\dd{y} \\
    = \exp\left(\frac{-a_l^2}{a_l^2\beta_j+4}(u+W_{pn}-\omega')^2\right) \int_{-\infty}^\infty\frac{1}{y}F\left(\frac{a_ly}{2}\right) \\
    \times\exp\left(-\frac{a_l^2\beta_j+4}{4\beta_j}\left(y-\frac{4(u+W_{pn}-\omega')}{a_l^2\beta_j+4}\right)^2\right)\dd{y}.
    \label{eq:dawson-gaussian-ex}
\end{multline}

The centre of the Gaussian function in the integral is far from the centre of the Dawson function. Since the Gaussian function decays quickly, it will pick out the value of the Dawson function near the centre of the Gaussian, $y=4(u+W_{pn}-\omega')/(a_l^2\beta_j+4)$. Therefore, we can approximate this integral as the product of the Dawson function evaluated at the centre of the Gaussian and a Hilbert transform of the Gaussian. Thus, the terms of this form in the integral over $y$ result in the product of a Gaussian function with two Dawson functions.

$\hat{S}_{pq}(\omega,\omega',\epsilon_j)$ is then a single integral over 49 terms that depend on $u$. However, we are only considering up to second order effects, which means we can neglect any terms of third order or higher in the polarizability. This leaves seventeen terms to consider.

Seven of the terms are integrals of a Dirac delta function multiplied by the result of the integral over $y$. Two of the terms are Hilbert transforms of Gaussian functions, which integrate to Dawson functions as discussed previously. The other eight terms can be integrated and simplified in the following way:

\begin{enumerate}
    \item Use partial fraction decomposition to separate the term into two terms with poles at $u=0$ and $u=\omega-W_{pn}$.
    \item Both terms take the form of a Gaussian function centred at $\omega'$ multiplied by one or two Dawson function and divided by $W_{pn}-\omega$.
    \item Since the width of the Gaussian function is much narrower than the difference between $W_{pn}$ and $\omega'$, we can approximate the denominators as $W_{pn}-\omega'$.
    \item The terms integrated over a pole at $u=0$ have a Dawson function with an input on the order of the pulse width times the laser frequency. Since we are in the adiabatic regime, this  is very large and the Dawson function $F(x)$ can be approximated as $1/(2x)$.
    \item The terms integrated over a pole at $u=\omega-W_{pn}$ have only Dawson functions that depend on $\omega$ as $\omega-\omega'$. Since these are multiplied by a Gaussian function centred at $\omega'$, the input to these Dawson functions is therefore small and we can approximate the Dawson function $F(x)$ as $x$.
\end{enumerate}

After making these approximations, we can take the inverse Fourier transform of $\hat{S}_{pq}(\omega,\omega',\epsilon_j)$ and return to the time domain. The only dependence on $W_{pn}$ remaining in $S_{pq}(t,\omega',\epsilon_j)$ is in the denominator of the terms. The denominators are either $W_{pn}^2-\omega'^2$ from the terms with the pole at $u=0$ or $W_{pn}-\omega'$ from all the others.

We can now consider the entire equation for $\expval{p}$,

\begin{multline}
    \expval{p}=\sum_{n=3}^\infty\sum_j\sum_{p=1,2}\sum_{q=1,2}p_{pn}p_{nq} \\
    \times\left(S_{pq}(t,\omega_{j+p-q},\epsilon_j)+S_{pq}(t,-\omega_{j-p+q},\epsilon_j^*)\right)+c.c.
    \label{eq:dipole-moment-sum}
\end{multline}

We eliminate the terms where $p=q$ and consider the sum over $n$ of $S_{pq}(t,\omega_{j+p-q},\epsilon_j)$ and $S^*_{qp}(t,-\omega_{j-p+q},\epsilon_j^*)$. The terms with $W_{pn}^2-\omega'^2$ in the denominator are multiplied by $i$ and sum to zero while the terms with $W_{pn}-\omega'$ in the denominator sum to the polarizability $\alpha_{12}$.

Thus we obtain the polarization to second order as the following simplified equation,

\begin{multline}
    \expval{p}=\frac{\alpha_{12}}{2}\sum_j\left[E_je^{i\phi_j}e^{i\omega_{j-1}t}R_{12}(t,\beta_j)\right. \\
    +\left.E_j^*e^{-i\phi_j}e^{-i\omega_{j+1}t}R_{12}(t,\beta_j^*)+c.c.\right]
    \label{eq:expval-dipole-moment}
\end{multline}

\noindent where $R_{12}$ are given in Appendix \ref{app:polarization}. 

To better understand this equation, we will assume that the Raman orders maintain approximately the same width and chirp, which we will denote $a$ and $b$ respectively. We define also the parameter $r\equiv a^2b$, which is a ratio of the real and imaginary parts of $\beta$ and the sum of electric field amplitudes $\mathcal{I}\equiv\sum_j|E_j|^2$. We then have the polarization to second order is,

\begin{multline}
    \expval{p}=\frac{\alpha_{12}}{2}\sum_j\left[E_je^{i\phi_j}e^{i\omega_{j-1}t}e^{ibt^2/4}R_{12}(t,r)\right. \\
    +\left.E_j^*e^{-i\phi_j}e^{-i\omega_{j+1}t}e^{-ibt^2/4}R_{12}(t,-r)+c.c.\right]
    \label{eq:polarization}
\end{multline}

\noindent where

\begin{widetext}
\begin{multline}
    {R}_{12}(t,r)=\sum_jE_j^*E_{j-1}e^{-i(\phi_j-\phi_{j-1})}\left\{\left[\frac{\pi a^2\alpha_{12}}{64\hbar^2}\left(3\alpha_{11}+\alpha_{22}\right)\mathcal{I}+ \frac{ia\alpha_{12}}{4\hbar}\sqrt{\frac{\pi}{2}}\right]e^{-t^2/(4a^2)} \right.\\
    +\sqrt{\frac{2}{\pi}}\frac{3-ir}{1-ir}\left[\frac{\pi a^2\alpha_{12}}{32\hbar^2}\left(3\alpha_{11}+\alpha_{22}\right)\mathcal{I}+\frac{ia\alpha_{12}}{4\hbar}\sqrt{\frac{\pi}{2}}\right]\frac{t}{a}e^{-3t^2/(4a^2)} \\
    +\left.\frac{1}{1-ir}\frac{a^2\alpha_{12}}{16\hbar^2}\left(3\alpha_{11}+2\alpha_{22}\right)\mathcal{I}\left(\frac{t^2}{a^2}(5-ir)-2\right)e^{-5t^2/(4a^2)}\right\}.
    \label{eq:polarization-term}
\end{multline}
\end{widetext}

At zeroth order, the time-dependence of the polarization is the same as the perturbing electric field as expected. However, the higher order terms introduce a more complex time-dependence that is narrower in time and depends on higher powers of time. We expect this trend to continue if we consider higher order effects, which would introduce for example a term proportional to $t^3/a^3e^{-7t^2/(4a^2)}$ at third order.

%% file: spectrum.tex
The polarization given in the previous section is the response of the potential in time. The frequencies emitted by the Raman medium are therefore given by the magnitude of the Fourier transform of the polarization. By taking the Fourier transform of Eq. \ref{eq:polarization}, we obtain the MRG spectrum, $W(\omega)=|\mathcal{F}\{\expval{p}\}|^2$.

\begin{multline}
    W(\omega)=\left|\frac{\alpha_{12}}{2}\sum_j\left[E_je^{i\phi_j}\hat{R}_{12}(\omega-\omega_{j-1},r)\right.\right. \\
    + E_j^*e^{-i\phi_j}\hat{R}_{12}(\omega+\omega_{j+1},-r)
    + E_j^*e^{-i\phi_j}\hat{R}_{12}^*(\omega+\omega_{j-1},r) \\
    +\left.\left.E_je^{i\phi_j}\hat{R}_{12}^*(\omega-\omega_{j+1},-r)\right]\right|^2
    \label{eq:W}
\end{multline}

\noindent where

\begin{widetext}
\begin{multline}
    {R}_{12}(\nu,r)=\sum_jE_j^*E_{j-1}e^{-i(\phi_j-\phi_{j-1})}\left\{\frac{a\sqrt{\pi}}{\sqrt{1-ir}}\left[\frac{\pi a^2\alpha_{12}}{32\hbar^2}\left(3\alpha_{11}+\alpha_{22}\right)\mathcal{I}+ \frac{ia\alpha_{12}}{2\hbar}\sqrt{\frac{\pi}{2}}\right]\exp\left(-\frac{a^2\nu^2}{1-ir}\right) \right.\\
    -\frac{ia^2\sqrt{2}}{(1-ir)\sqrt{3-ir}}\left[\frac{\pi a^2\alpha_{12}}{8\hbar^2}\left(3\alpha_{11}+\alpha_{22}\right)\mathcal{I}+\frac{ia\alpha_{12}}{\hbar}\sqrt{\frac{\pi}{2}}\right]\nu\exp\left(-\frac{a^2\nu^2}{3-ir}\right) \\
    -\left.\frac{a^3\sqrt{\pi}}{(1-ir)(5-ir)^{3/2}}\frac{a^2\alpha_{12}}{2\hbar^2}\left(3\alpha_{11}+2\alpha_{22}\right)\mathcal{I}\nu^2\exp\left(-\frac{a^2\nu^2}{5-ir}\right)\right\}.
    \label{eq:W-term}
\end{multline}
\end{widetext}

We can specifically find the spectrum radiated by the Raman medium due to the initial excitation. Initially, the only electric fields are the pump and the probe beams with frequency $\omega_0$ and $\omega_{-1}$ and amplitudes $E_0$ and $E_{-1}$ respectively. We will assume the electric field amplitudes are the same since we will compare to experiments where this is approximately the case. 

\begin{figure}
    \includegraphics[width=\columnwidth]{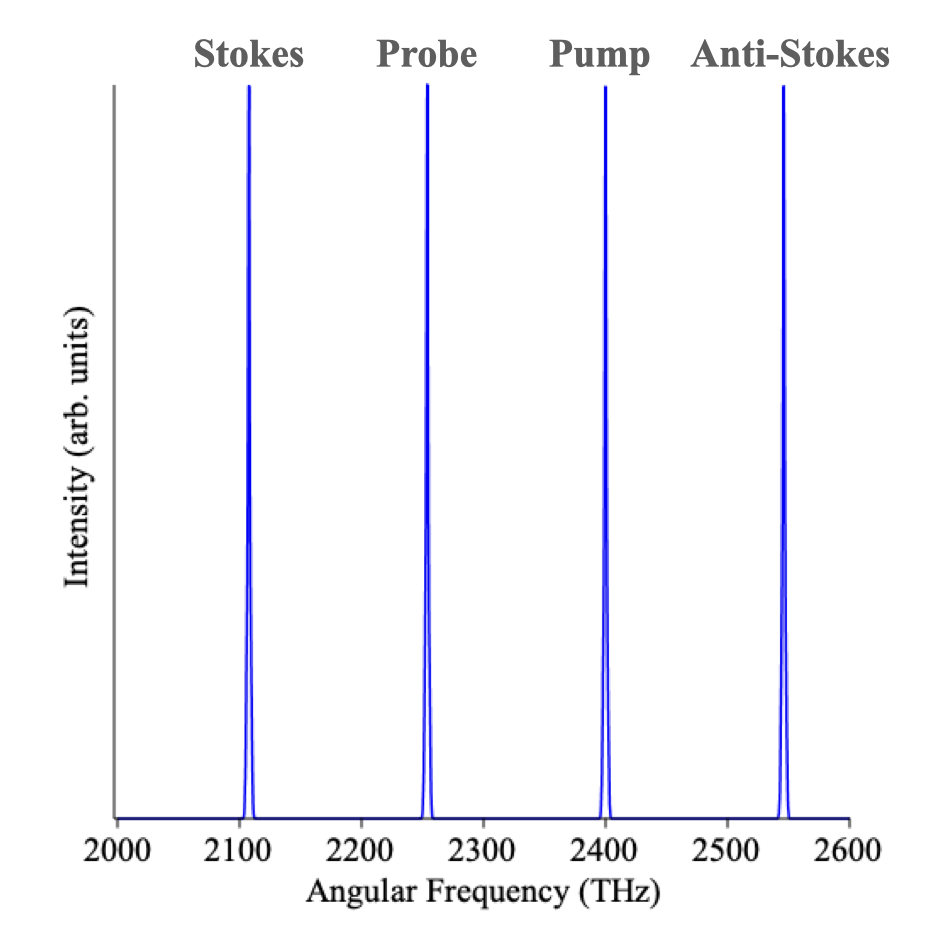}
    \caption{Spectrum radiated by Raman medium upon excitation by just the pump and probe electric fields}
    \label{fig:full-initial-spectrum}
\end{figure}

In Fig. \ref{fig:full-initial-spectrum}, we see that the spectrum has four peaks. The two central peaks correspond to the pump and probe frequencies while the lower and higher frequency peaks correspond to the first Stokes and anti-Stokes peaks respectively. 

For the rest of this paper, we will be inspecting the behaviour of the first anti-Stokes peak (near $\omega_1=\omega_0+\omega_R$). The dependence of the first anti-Stokes peak on varying electric field amplitudes and chirps are shown below.

\begin{figure*}
    \centering
    \includegraphics[width=\textwidth]{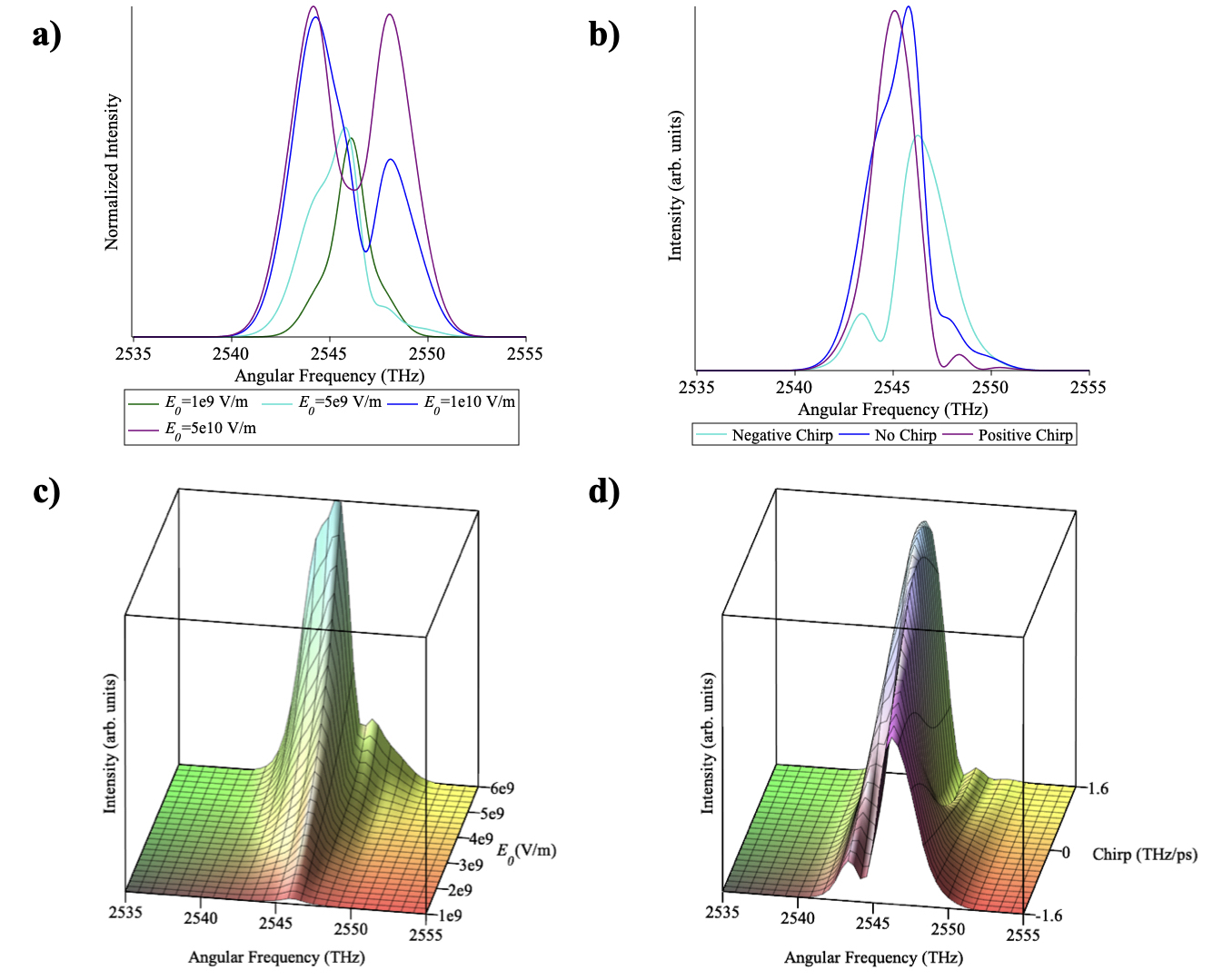}
    \caption{Spectrum of radiation emitted by Raman medium upon initial excitation around the anti-Stokes frequency with varying (a) electric field amplitudes $E_0=E_{-1}$ ($b=0$) and (b) chirp $b$ ($E_0=E_{-1}=5\times10^9$V/m). Plots (c) and (d) show the continuous change in spectrum with (c) electric field amplitude and (d) chirp. We have used the example values $\omega_0=2\pi382$THz, $\omega_R=2\pi23.25$THz, $a=800$fs,  $\alpha_{12}=1.465\times10^{-42}$Cm$^2$/V, $\alpha_{11}=1.456\times10^{-42}$Cm$^2$/V, and $\alpha_{22}=1.478\times10^{-42}$Cm$^2$/V and assumed $E_0=E_{-1}$ \cite{Abdullah2018, Rand2022}}
    \label{fig:spectrum_plots}
\end{figure*}

%% file: discussion.tex
The energies in Fig. \ref{fig:spectrum_plots} correspond to the electric field amplitude expected from laser pulses with mJ order energy that lose several orders of magnitude to processes besides Raman. 

At low energies and no chirp ($b=0$), the spectrum $W(\omega)$ (Eq. \ref{eq:W}) near the anti-Stokes frequency is a singular peak centred at the anti-Stokes frequency. As we increase electric field amplitude (and thereby energy), Figs. \ref{fig:spectrum_plots}a and \ref{fig:spectrum_plots}c show the main peak shifting to the red as the spectrum splits and the secondary peak slowly grows. The red-shifting of the central peak with increasing energy agrees with experiment \cite{Yan2013}. If we could increase the energy by an order of magnitude in a similar experiment, we would expect to clearly see the secondary peak. This provides an avenue to test our theoretical model.

When we add chirp to the pulses, we introduce another asymmetry. Negative chirps and positive chirps produce sidebands shifted to the red and blue respectively. However, we can see from Figs. \ref{fig:spectrum_plots}b and \ref{fig:spectrum_plots}d that the positively chirped sideband is much smaller than the negatively chirped sideband and would likely be lost in experimental noise. This occurs because the electric field amplitude asymmetry and the positive chirp asymmetry both suppress the same side of the double peak. The spectrum for MRG with negatively chirped pulses has a red-shifted shoulders, as observed experimentally \cite{Turner2008, Abdullah2018,Zujun2021}. This theory indicates that the red-shifted shoulders in MRG occur because the asymmetry from the chirp offsets the asymmetry in how the secondary peak emerges with increasing electric field amplitude.

%% file: conclusion.tex
We have presented a theory for the analytic spectrum of radiation emitted by a Raman medium excited by a multiwave electric field. We predict that increasing the energy of the pump and probe lasers will cause the anti-Stokes spectrum to red-shift until a double-peaked structure emerges at high energies. The former has been observed experimentally and the later provides a potential test for the model used to derive the analytic spectrum. 

We also predict sidebands to the red and blue for negatively and positively chirped fields respectively. The blue sideband has negligible intensity while the red sideband is clearly visible and resembles the red-shifted shoulder observed in experimental MRG.

%% file: appendix.tex
\section{Dawson Function \label{app:dawson}}

\begin{figure}
    \includegraphics[width=\columnwidth]{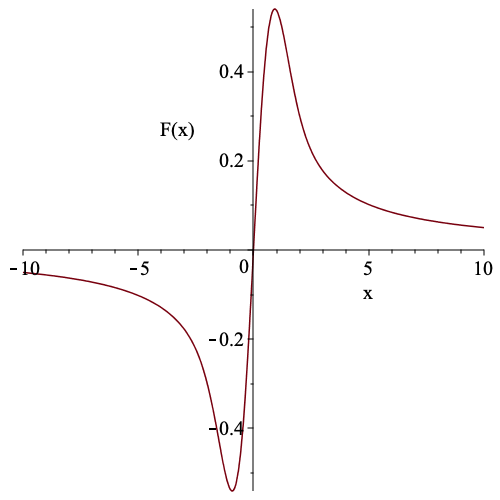}
    \caption{Dawson function}
    \label{fig:dawson}
\end{figure}

The Dawson function is defined as \cite[Eq.~7.2.5]{NIST:DLMF}:

\begin{equation}
    F(z)=e^{-z^2}\int_0^ze^{t^2}\dd{t}
    \label{eq:dawson}
\end{equation}

\noindent and can be graphed for real $z$ as in Fig. \ref{fig:dawson}.

The following proof is a complex version of that given by \cite{WillG2021}.

The Hilbert transform of a Gaussian function is given by Eq. \ref{eq:hilbert-transform}. We first make the substitution $u=y-x$.

\begin{equation}
    H(y)=\frac{1}{\pi}P.V.\int_{-\infty}^\infty\frac{e^{-(y-u)^2}}{u}(-\dd{u}).
    \label{eq:hilbert-transform-usub}
\end{equation}

We can then write out the principle value explicitly,

\begin{equation}
    H(y)=\frac{-1}{\pi}\lim_{\epsilon\to0}\left(\int_{-\infty}^{-\epsilon}\frac{e^{-(y-u)^2}}{u}\dd{u}+\int_{\epsilon}^\infty\frac{e^{-(y-u)^2}}{u}\dd{u}\right).
    \label{eq:H-pv-explicit}
\end{equation}

We now make the substitution $z=-u$ in the first integral and reverse the order of integration to simplify $H(y)$ to,

\begin{equation}
    H(y)=\frac{2}{\pi}e^{-y^2}\lim_{\epsilon\to0}\int_\epsilon^\infty\frac{\sinh(2yz)}{z}e^{-z^2}\dd{z}.
    \label{eq:H-sinh-form}
\end{equation}

We now have an integrand that is a holomorphic function of $y$ and is continuous in $z$ over the bounds of integration. Therefore, we can differentiate the integral in $H(y)$,

\begin{equation}
    \frac{\partial}{\partial y}\int_\epsilon^\infty\frac{\sinh(2yz)}{z}e^{-z^2}\dd{z}=2\int_\epsilon^\infty\cosh(2yz)e^{-z^2}\dd{z},
    \label{eq:H-integral-derivative}
\end{equation}

which is now just an integral of complex Gaussian functions that we can easily integrate to $\sqrt{\pi}e^{y^2}$.

Since $\sinh(0)=0$, we can therefore write the Hilbert transform of a Gaussian in the simplified integral form that can be identified with the definition of the Dawson function above.

\begin{equation}
    H(y)=\frac{2}{\sqrt{\pi}}e^{-y^2}\int_0^ye^{t^2}\dd{t}=\frac{2}{\sqrt{\pi}}F(y)
    \label{eq:H-F-equivalence}
\end{equation}

\section{Polarization with Differing Pulse Shapes \label{app:polarization}}

If we do not make the approximation that all Raman order electric fields have the same width and chirp, we instead write the polarization (Eq. \ref{eq:exp-dipole-moment}) as a sum of unsimplified $R_{12}(t,\beta_j)$.

\begin{widetext}
\begin{multline}
    {R}_{12}(t, \beta_j)=\eta_1^{(0)*}\eta_2^{(0)}e^{-\beta_jt^2/4} \\
    +\sum_l\left[\sqrt{\frac{2}{\pi}}\frac{a_l^2\beta_j+2}{a_l^2\beta_j}\left[\eta_{1a}^{(1)*}(l)\eta_2^{(0)}+\eta_1^{(0)*}\eta_{2a}^{(1)}(l)\right]\frac{t}{a_l}e^{-(a_l^2\beta_j+2)t^2/(4a_l^2)}\right. \\
    +\frac{1}{\sqrt{\pi}\beta_j}\left[(\beta_j+\beta_l^*+\beta_{l-1})\eta_{1b}^{(1)*}(l)\eta_2^{(0)}\sqrt{\beta_l^*+\beta_{l-1}}te^{-t^2(\beta_j+\beta_l^*+\beta_{l-1})/4}\right. \\
    \left.+(\beta_j+\beta_l+\beta_{l-1}^*)\eta_1^{(0)*}\eta_{2b}^{(1)}(l)\sqrt{\beta_l+\beta_{l-1}^*}te^{-t^2(\beta_j+\beta_l+\beta_{l-1}^*)/4}\right] \\
    +\frac{8}{\pi\beta_ja_l^2}\left[\eta_{1a}^{(2)*}(l)\eta_2^{(0)}+\eta_1^{(0)*}\eta_{2a}^{(2)}(l)\right]\left(\frac{t^2}{a_l^2}(a_l^2\beta_j+4)-2\right)e^{-(a_l^2\beta_j+4)t^2/(4a_l^2)} \\
    +\frac{4}{\pi\beta_j}\left[(\beta_l^*+\beta_{l-1})\eta_{1b}^{(2)*}(l)\eta_2^{(0)}(t^2(\beta_j+2\beta_l^*+2\beta_{l-1})-2)e^{-(\beta_j+2\beta_l^*+2\beta_{l-1})t^2/4} \right. \\
    \left.+(\beta_l+\beta_{l-1}^*)\eta_1^{(0)*}\eta_{2b}^{(2)}(l)(t^2(\beta_j+2\beta_l+2\beta_{l-1}^*)-2)e^{-(\beta_j+2\beta_l+2\beta_{l-1}^*)t^2/4} \right] \\
    + \sum_k\left[\frac{2}{\pi a_ka_l\beta_j}\eta_{1a}^{(1)*}(l)\eta_{2a}^{(1)}(k)\left(t^2\frac{(a_l^2\beta_j+2)a_k^2+2a_l^2}{a_k^2a_l^2}-2\right)e^{-\frac{(a_l^2\beta_j+2)a_k^2+2a_l^2}{4a_k^2a_l^2}t^2}\right. \\
    +\frac{\sqrt{2}}{\pi\beta_j}\left[\sqrt{\frac{\beta_k+\beta_{k-1}^*}{a_l^2}}\eta_{1a}^{(1)*}(l)\eta_{2b}^{(1)}(k)\left(t^2\frac{a_l^2(\beta_j+\beta_k+\beta_{k-1}^*)+2}{a_l^2}-2\right)e^{-(a_l^2(\beta_j+\beta_k+\beta_{k-1}^*)+2)t^2/(4a_l^2)}\right. \\
    +\left.\sqrt{\frac{\beta_k^*+\beta_{k-1}}{a_l^2}}\eta_{1b}^{(1)*}(k)\eta_{2a}^{(1)}(l)\left(t^2\frac{a_l^2(\beta_j+\beta_k^*+\beta_{k-1})+2}{a_l^2}-2\right)e^{-(a_l^2(\beta_j+\beta_k^*+\beta_{k-1})+2)t^2/(4a_l^2)}\right] \\
    +\frac{\sqrt{(\beta_k^*+\beta_{k-1})(\beta_l+\beta_{l-1}^*)}}{\pi\beta_j}\eta_{1b}^{(1)*}(k)\eta_{2b}^{(1)}(l)\left((\beta_j+\beta_k^*+\beta_{k-1}+\beta_l+\beta_{l-1}^*)t^2-2\right) \\
    \times e^{-(\beta_j+\beta_k^*+\beta_{k-1}+\beta_l+\beta_{l-1}^*)t^2/4}+\left[\eta_{1a}^{(3)*}(l)\eta_2^{(0)}+\eta_1^{(0)*}\eta_{2a}^{(3)}(l)\right]e^{-(a_l^2\beta_j+4)t^2/(4a_l^2)} \\
    +\left[\eta_{1b}^{(3)*}(l)\eta_2^{(0)}e^{-(\beta_j+2\beta_l^*+2\beta_{l-1})t^2/4}+\eta_1^{(0)*}\eta_{2b}^{(3)}(l)e^{-(\beta_j+2\beta_l+2\beta_{l-1}^*)t^2/4} \right]
    \label{eq:full-polarization-term}
\end{multline}
\end{widetext}

\noindent where we have the time independent coefficients $\eta$ from the amplitudes $c_1(t)$ and $c_2(t)$,

\begin{align}
\begin{split}
    \eta_1^{(0)*}\eta_2^{(0)}&=\sum_l\xi_{2b}^{(1)}(l)+\sum_l\sum_k\left[\xi_{2a}^{(2)}(l,k)\frac{\sqrt{2}\pi a_l}{4\sqrt{\beta_k^*+\beta_{k-1}}}\right. \\
    &+\frac{\alpha_{11}}{\alpha_{22}}\xi_{2a}^{(2)}(k,l)\frac{\sqrt{2}\pi a_k}{4\sqrt{\beta_l^*+\beta_{l-1}}}+\xi_{1a}^{(1)*}(l)\xi_{2b}^{(1)}(k) \\
    \eta_{1a}^{(1)*}(l)\eta_2^{(0)}&=\sum_k\xi_{1a}^{(1)*}\xi_{2b}^{(1)}(k) \\
    \eta_{1b}^{(1)*}(l)\eta_2^{(0)}&= \eta_{1a}^{(2)*}(l)\eta_2^{(0)} = \eta_{1b}^{(2)*}(l)\eta_2^{(0)} \\
    &=\eta_{1a}^{(3)*}(l)\eta_2^{(0)}=\eta_{1b}^{(3)*}(l)\eta_2^{(0)}=0 \\
    \eta_1^{(0)*}\eta_{2a}^{(1)}(l)&=\sum_k\xi_{2a}^{(2)}(l,k)\frac{\sqrt{2}\pi a_l}{2\sqrt{\beta_k^*+\beta_{k-1}}} \\
    \eta_1^{(0)*}\eta_{2b}^{(1)}(l)&=\xi_{2b}^{(1)}(l) + \sum_k\left[\xi_{1a}^{(1)*}(k)\xi_{2b}^{(1)}(l)\right.\\
    &+\left.\frac{\alpha_{11}}{\alpha_{22}}\xi_{2a}^{(2)}(k,l)\frac{\sqrt{2}\pi a_k}{2\sqrt{\beta_l^*+\beta_{l-1}}}\right] \\
    \eta_1^{(0)*}\eta_{2a}^{(2)}(l)&=\sum_k\xi_{2a}^{(2)}(l,k)\frac{\sqrt{2}\pi a_l}{4\sqrt{\beta_k^*+\beta_{k-1}}} \\
    \eta_1^{(0)*}\eta_{2b}^{(2)}(l)&=\sum_k\frac{\alpha_{11}}{\alpha_{22}}\xi_{2a}^{(2)}(k,l)\frac{\sqrt{2}\pi a_k}{4\sqrt{\beta_l^*+\beta_{l-1}}} \\
    \eta_1^{(0)*}\eta_{2a}^{(3)}(l)&=\sum_k\xi_{2a}^{(2)}(l,k)\left(\frac{1}{\beta_k^*+\beta_{k-1}}-\frac{a_l/\sqrt{2}}{\sqrt{\beta_k^*+\beta_{k-1}}}\right) \\
    \eta_1^{(0)*}\eta_{2b}^{(3)}(l)&=\sum_k\frac{\alpha_{11}}{\alpha_{22}}a_k\xi_{2a}^{(2)}(k,l)\left(\frac{a_k}{2}-\frac{1}{\sqrt{2}\sqrt{\beta_l^*+\beta_{l-1}}}\right) \\
    \eta_{1a}^{(1)*}(l)\eta_{2a}^{(1)}(k)&=\eta_{1b}^{(1)*}(k)\eta_{2a}^{(1)}(l)=\eta_{1b}^{(1)*}(k)\eta_{2b}^{(1)}(l)=0 \\
    \eta_{1a}^{(1)*}(l)\eta_{2b}^{(1)}(k)&=\xi_{1a}^{(1)*}(l)\xi_{2b}^{(1)}(k)
\end{split}
\end{align}

\noindent and

\begin{align}
    \begin{split}
        \xi_{1a}^{(1)}(l)&=\frac{i\alpha_{11}}{4\hbar}|A_l|^2\sqrt{\frac{\pi}{2}}a_l \\
        \xi_{1a}^{(2)}(l,k)&=\frac{\alpha_{11}^2}{16\hbar^2}|A_l|^2|A_k|^2 \\
        \xi_{1b}^{(2)}(l,k)&=\frac{\alpha_{12}^2}{16\hbar^2}A_lA_{l-1}^*A_k^*A_{k-1} \\
        \xi_{2b}^{(1)}(l)&=\frac{i\alpha_{12}}{4\hbar}A_l^*A_{l-1}\sqrt{\frac{\pi}{\beta_l^*+\beta_{l-1}}} \\
        \xi_{2a}^{(2)}(l,k)&=\frac{\alpha_{12}\alpha_{22}}{16\hbar^2}|A_l|^2A_k^*A_{k-1}
    \end{split}
\end{align}

We have again only considered the coefficients $\eta$ to second order in the polarizability.